# Engineered second-order nonlinearity in silicon nitride

YI ZHANG,[1] JUNIYALI NAURIYAL,[2] MEITING SONG,[1] MARISSA GRANADOS BAEZ,[1] XIAOTONG HE,[1] TIMOTHY MACDONALD[3], AND JAIME CARDENAS[1*]

[1]*The Institute of Optics, University of Rochester, Rochester, NY 14627, USA*
[2]*Department of Electrical and Computer Engineering, University of Rochester, Rochester, NY 14627, USA*
[3]*Department of Electrical and Computer Engineering, California State University, Northridge, Northridge, CA 91330, USA*

*\*jaime.cardenas@rochester.edu*

**Abstract:** The lack of a bulk second-order nonlinearity ($\chi^{(2)}$) in silicon nitride ($Si_3N_4$) keeps this low-loss, CMOS-compatible platform from key active functions such as Pockels electro-optic (EO) modulation and efficient second harmonic generation (SHG). We demonstrate a successful induction of $\chi^{(2)}$ in $Si_3N_4$ through electrical poling with an externally-applied field to align the Si-N bonds. This alignment breaks the centrosymmetry of $Si_3N_4$, and enables the bulk $\chi^{(2)}$. The sample is heated to over 500°C to facilitate the poling. The comparison between the EO responses of poled and non-poled $Si_3N_4$, measured using a $Si_3N_4$ micro-ring modulator, shows at least a 25X enhancement in the $r_{33}$ EO component. The maximum $\chi^{(2)}$ we obtain through poling is 0.24pm/V. We observe a remarkable improvement in the speed of the measured EO responses from 3GHz to 15GHz (3dB bandwidth) after the poling, which confirms the $\chi^{(2)}$ nature of the EO response induced by poling. This work paves the way for high-speed active functions on the $Si_3N_4$ platform.

## 1. Introduction

Silicon nitride ($Si_3N_4$) is a high-performance platform for versatile on-chip photonic devices [1-4] because of its low propagation loss, broad transparency window (400-6700nm [5]) and good compatibility with complementary metal-oxide semiconductor (CMOS) processing. However, $Si_3N_4$ lacks an intrinsic second-order nonlinearity ($\chi^{(2)}$) due to its centrosymmetric structure [6], which limits its applications from some critical active functions, such as electro-optic (EO) modulation and second harmonic generation (SHG).

Previous works have made great progress in realizing $\chi^{(2)}$-related phenomena on a silicon nitride platform. Pioneer studies report a SHG signal that originates from symmetry breaking at the interface of a multilayer stack of amorphous $Si_{1-x}N_x$:H films with various compositions [7]. An electro-optic response is also observed in a stacked $Si_3N_4$ multilayer [8]. Other works point out the existence of a bulk second-order nonlinearity in deposited $Si_3N_4$ films [9-10]; possible explanations include the presence of silicon nanocrystals [9] and a built-in static electric field [10] that leads to an electric field induced second harmonic generation (EFISHG). There are reports of SHG in Si-rich silicon nitride [11-12]. The use of high-quality micro-ring resonators [13] and nano gratings [14], in which the optical field at the pump frequency is amplified, further improves the efficiency of SHG in silicon nitride. Recently, an all-optical poling (AOP) method [15-16] has demonstrated record-breaking SHG [17] and sum-frequency conversion (SFC) [18] in $Si_3N_4$. The optical poling process in this approach automatically forms quasi-phase matching (QPM) gratings in the $Si_3N_4$ waveguide [15-20] and the QPM condition can be tuned by re-poling the device at the corresponding wavelength [19-20]. On the other hand, to realize Pockels EO modulation on the $Si_3N_4$ platform, other works have explored heterogeneous integration of other materials with a usable EO response, such as lithium niobate (LN) [21], barium titanate (BTO) [22], lead zirconate titanate (PZT) [23], zinc oxide and zinc sulfide [24], EO polymers [25], and two-dimensional (2D) materials [26-28], onto the $Si_3N_4$ platform. These approaches, however, complicate the fabrication process, introduce extra loss and/or are not CMOS-compatible. Alternatively, a very recent work by Zabelich et. al. [29] demonstrates an electro-optic response of tens of kHz on a pure $Si_3N_4$ platform induced by electrical poling at elevated temperatures.

In this paper, we propose and demonstrate an induction of a second-order nonlinearity for electro-optic modulation in silicon nitride through electrically poling the film and aligning the Si-N bonds. We show an enhancement of at least 25X of the EO response in a $Si_3N_4$ micro-ring resonator. The 3dB bandwidth of the EO response after poling improves from 3GHz to at least 15GHz. Further analysis supports that this enhancement directly derives from the induction of $\chi^{(2)}$ in $Si_3N_4$. Khurgin et al. [30] hypothesized that the Si-N bonds in $Si_3N_4$ possess a second-order hyperpolarizability comparable to the Ga-As bonds in gallium arsenide (GaAs), whose $\chi^{(2)}$ is over 300pm/V [30]. Once the centrosymmetry in $Si_3N_4$ is broken, e.g., through electrical poling, a non-trivial bulk second-order nonlinearity will appear and thus enable the Pockels electro-optic effect. Electrical poling has been used to engineer material properties, such as to periodically pole various nonlinear crystals for QPM [6] and to induce a second order nonlinearity in polymers [31], silica glass [32], optical fibers [33], and very recently in $Si_3N_4$ [29].

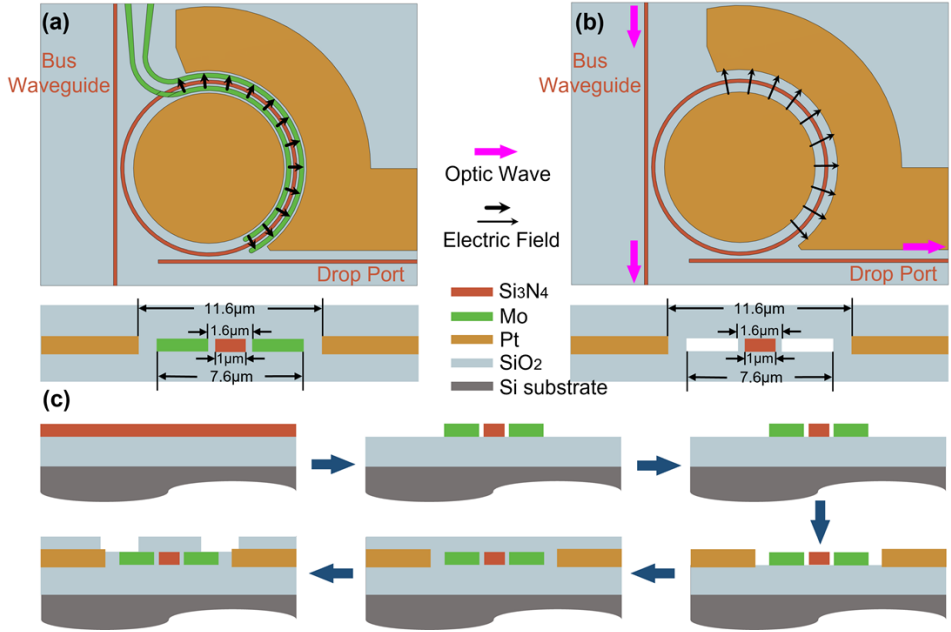

Fig. 1. Top view (top cladding hidden, not to scale) and typical cross section (not to scale) of the device (a) before and (b) after removal of Mo electrodes. A voltage is applied between the Mo and the Pt electrode pair for poling and high-speed modulation, respectively. (c) Fabrication process of the device.

## 2. Device design and fabrication

We pole silicon nitride by embedding a removable set of electrodes into the device. We deposit a pair of Molybdenum (Mo) electrodes sandwiching the $Si_3N_4$ ring (Figure 1(a)). The gap between the electrodes and the ring waveguide is 300nm (edge to edge). The Mo electrodes generate an average horizontal electric field of $4.23 \times 10^3$V/cm per volt of bias applied in the $Si_3N_4$ waveguide (simulated with COMSOL Multiphysics®). We apply 160V across the Mo electrode pair and build a poling field of 0.676MV/cm. We choose Mo as the material for the poling electrodes because it is compatible with the high temperatures reached during the poling process (See Section 3) and it can be removed (Fig.1(b)) using xenon difluoride ($XeF_2$) etch to eliminate the loss it introduces to the ring resonator. The $XeF_2$ etch has a high selectivity to other materials used in the device and thus, it does not influence the poling results. We deposit another set of electrodes made of platinum (Pt) outside the Mo ones for characterizing the electro-optic response of the device after removal of the sacrificial Mo electrodes (Fig.1(b)). To fabricate the device (Figure 1(c)), we deposit 300nm of $Si_3N_4$ using low pressure chemical vapor deposition (LPCVD) over 4 μm of thermally-grown $SiO_2$ on a 4-inch Si wafer. We define the waveguide and the ring resonator using electron-beam lithography and etch the structure with inductively coupled plasma reactive-ion etching (ICP-RIE). To create the poling electrodes, we pattern them with deep ultraviolet (DUV) photolithography and sputter and lift-off 300nm of Mo. 400nm Pt is then deposited in the same way. Afterwards, we deposit a cladding layer of 2.3μm of $SiO_2$ using plasma-enhanced chemical vapor deposition (PECVD) on top of the device and open vias using DUV photolithography and ICP-RIE to expose the Pt electrodes and later removal of the Mo electrodes.

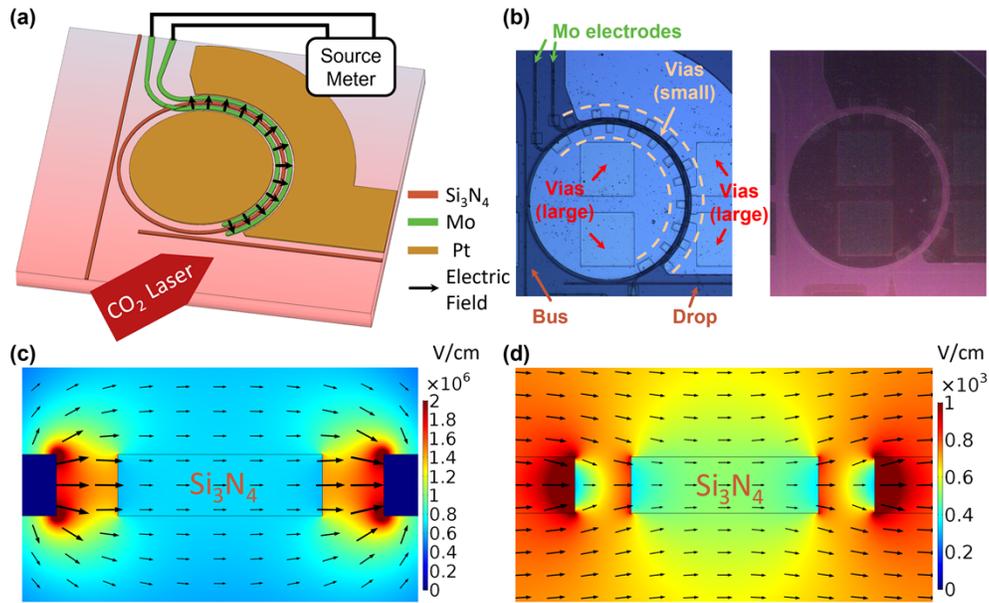

Fig 2. (a) Schematic (top cladding hidden, not to scale) of poling the device at high temperature. (b) Microscope image of the device under room (left) and high (right) temperature. External illumination is turned off when taking the high-temperature image and the red color comes from thermal radiation of the chip. (c-d) Simulated (COMSOL Multiphysics®) electric field profile in the waveguide (c) when applying 160V bias to pole the device and (d) when applying a unitary volt for high-speed modulation. Note the Mo electrodes in (c) are replaced with air in (d).

## 3. Device Poling

We pole the $Si_3N_4$ ring resonator by applying a strong electric field across it at high temperatures. A source meter (Keithley® 2470) generates the desired high voltage, and we apply it to the Mo electrodes through a pair of tungsten probes that is manipulated with micro-positioners. Before we enable the high voltage output, a continuous-wave (CW) $CO_2$ laser beam is focused onto the device, as shown in Fig. 2(a), to pre-heat it to a high temperature for more efficient poling [32]. We increase the laser power to approximately 10W and heat up the ring resonator to 550±40°C, which we calculate from the resistance change of the Pt electrodes as temperature varies. This value corresponds to the temperature at the middle of the heated ring resonator (Fig.2(a-b)). At higher temperatures it becomes challenging to maintain steady contact between the tungsten probes and the electrodes. We also observe that the arcing threshold drops as we increase the temperature, compromising the maximum poling field we can apply. The highest voltage we apply at this temperature is 160V, which generates a horizontal poling field of 0.676MV/cm in the $Si_3N_4$ waveguide (Fig.2(c)). We observe arcing between the exposed regions of the poling electrodes for a voltage of 180V or higher. The corresponding horizontal poling field at this threshold is 0.760MV/cm. The breakdown threshold of silicon nitride at room temperature is 3-8MV/cm [34]. The lower arcing threshold in our experiment is a result of high temperature as well as that part of the Mo electrodes are exposed thus arcing can happen through the heated air. The collateral vertical field generated in the ring resonator is at least two orders of magnitude smaller than the horizontal field except at the corners, (Fig.2(c)), so the poling of the $Si_3N_4$ resonator is mostly horizontal.

We enable the source meter output and pole the device for 5 minutes after it is heated up to the desired temperature. When the poling finishes, we turn off the heating laser immediately but maintain the poling field until the device dissipates its heat to the metal holder and cools down to room temperature (in approximately half a minute). The fast cooling while the poling field still exists helps to freeze the alignment of the poled Si-N bonds and minimizes their rollback to their original state. No significant damage or deformation is observed on the device during the poling process. The Mo electrodes used for poling are then removed using $XeF_2$ dry etch to eliminate the optical loss it introduces to the ring resonator. The $XeF_2$ etch is run at room temperature and has a high selectivity to $Si_3N_4$, $SiO_2$, and Pt, so it does not undermine the already-built alignment of the Si-N bonds.

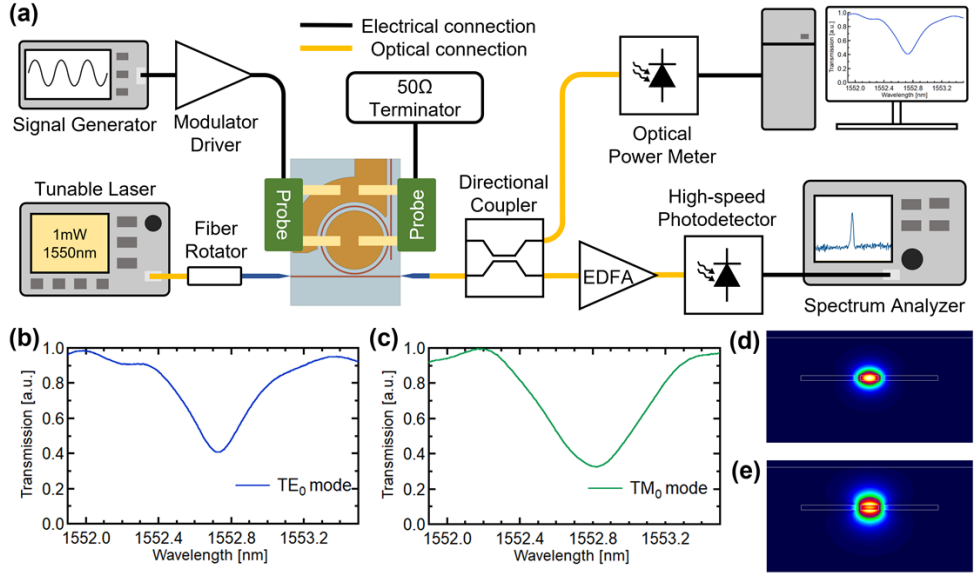

Fig 3. (a) Schematic of the setup used to measure the EO response of the poled Si3N4 device. (b-c) Typical transmission spectra of the device working in the (b) $TE_0$ and (c) $TM_0$ modes. (d-e) Simulated (FIMMWAVE, Photon Design®) electric field profile of the (b) $TE_0$ and (c) $TM_0$ modes after removal of the Mo electrodes.

## 4. Measurement of Electro-optic coefficient of silicon nitride

To examine the EO coefficient of $Si_3N_4$ and the impact of the poling process, we apply a radio frequency signal across the ring resonator with the remaining Pt electrodes and analyze the corresponding RF component in the output optical signal. We couple light into the device from a tunable laser (1mW) with a PM lensed fiber and collect the output optical power with a SM lensed fiber (Figure 3(a)). The ring resonator is designed to support the fundamental modes only. The transmission spectrum, monitored by an optical power meter, shows a typical loaded quality factor of 5,000 for the $TE_0$ mode and 3,000 for the $TM_0$ mode (Figure 3(b-c)) after removal of the Mo electrodes. The low quality factor results from the large loss at the rough sidewall $SiO_2$-air (Mo removed) interface, which is only 300nm away from the $Si_3N_4$ core, due to the isotropic nature of the $XeF_2$ etch. A fiber rotator at the input port enables launching TE or TM light into the device by rotating the PM fiber. The mode profiles calculated with the Finite Difference Method (FIMMWAVE, Photon Design®) in Fig.3(d) and (e) confirm considerable overlap between the mode and the interface. This loss can be reduced by optimizing the fabrication and the $XeF_2$ etching process.

We apply a modulation signal up to 15GHz. The sinusoidal signal is generated using a signal generator and amplified to up to approximately 8V through a modulator driver. The modulation signal is delivered to the Pt electrodes with a pair of high-speed (40GHz) probes and we terminate the transmission line with a 50Ω load to minimize signal reflection. We set the working wavelength of the laser at the slope of the resonance according to the transmission spectrum to optimize the EO response. The output of the photodetector is sent to a spectrum analyzer for characterizing the EO response of the device. Note that we amplify the output power to 10mW, the input limit of the photodetector, using a C-band Erbium-Doped Fiber Amplifier (EDFA) to make the high-frequency components more distinguishable from the background noise of the spectrum analyzer.

The RF-frequency electrical power $P_{SA}$ measured at the spectrum analyzer allows us to quantitatively characterize the EO coefficient of the $Si_3N_4$ in our device. The corresponding refractive index change of nitride $\Delta n_{SiN}$ can be determined through

$$P_{SA} = \eta \, (G \cdot ff \cdot \frac{dP}{d\lambda} \cdot \lambda \cdot \frac{\Gamma_{core} \Delta n_{SiN}}{n_g})^2 \tag{1}$$

where $G$ is the gain of the EDFA, $\Gamma_{core}$ is the confinement factor of the waveguide mode in the $Si_3N_4$ core, $n_g$ is the group index of the working mode, and $\lambda$ is the working wavelength. $dP/d\lambda$ is the slope of the transmission spectrum at the working wavelength and can be extracted from the spectrum data. $\eta$ is the optic-to-electric power conversion efficiency of the photodiode and we calibrate it to be 1.426±0.071μW/mW² in our test. $ff$ is the filling factor [35] that describes how efficiently the applied modulation field induces a resonance shift in a resonator with a certain geometry. In our case, since the deposited $Si_3N_4$ is amorphous and any asymmetry we introduce (alignment of bonds, waveguide bending, etc.) is radial, there is no mutual cancelling of resonance shift among different parts of the ring as in X-cut or Y-cut lithium niobate ring resonators [35]. Meanwhile, our device has the Mo electrodes and Pt electrodes sandwiching 44.4% of the ring, thus $ff$ is 44.4%.

The radial-only asymmetry we introduce into the initially amorphous $Si_3N_4$ turns it into a rotationally symmetric structure with one axis of symmetry. Such a structure is analogous to the point group $C_{\infty v}$ with its axis of symmetry (Z direction) along the (radial) direction of the poling field, leading to a linear electro-optic tensor with only five non-zero terms: $r_{13}=r_{23}$, $r_{33}$, and $r_{42}=r_{51}$ [6,36]. The electric field applied through the Pt electrode for the RF-speed test is also radial, thus the EO coefficients available to characterize in this device are $r_{13}$ and $r_{33}$, under the $TM_0$ mode and the $TE_0$ mode, respectively, using the equation [6]

$$\Delta n_{SiN_i} = -\frac{1}{2} r_{i3} n_{SiN_i}^3 E_{mod}, \, i=1,2,3 \tag{2}$$

where the amplitude of the applied RF modulation field $E_{mod}$ is proportional to the applied voltage $V_{mod}$ (both half peak-to-peak value), which we calibrate for all frequencies of interest. We expect a relatively negligible change of refractive index of the $Si_3N_4$ from the poling and therefore treat it as isotropic in our calculations. The average horizontal field across the $Si_3N_4$ core is $4.83 \times 10^2$V/cm per volt applied, according to simulations (Fig.2(d)), while the collateral vertical field is at least one order of magnitude smaller in the $Si_3N_4$ core (except at the corners). We ignore the effect of the collateral vertical field in the calculation of the EO coefficient. The high-speed probe loss vs frequency and the photodiode efficiency vs frequency are taken from the manufacturer specifications.

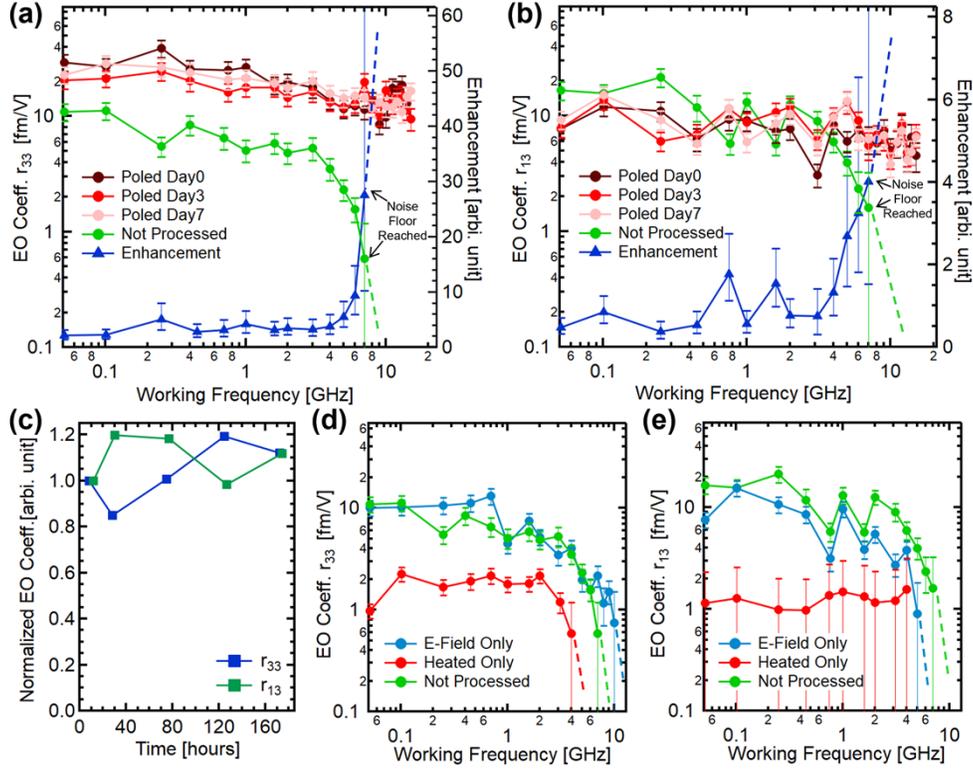

Fig 4. Measured EO coefficients $r_{33}$ ((a), (d)) and $r_{13}$ ((b), (e)) of the poled device (dark red, red, and pink curves in (a-b)), non-processed device (green curves), and devices processed only with high temperature (red curves in (d-e)) or strong electric field (blue curves in (d-e)). Dashed lines in (a-b, d-e) represent estimated trend lines of corresponding values at higher frequencies. From the beginning of these dashed lines, the response in the spectrum analyzer could not be distinguished from the background noise. Therefore, the lower error bars of these points go all the way to zero (EO coefficients) or infinity (enhancement). Performance of the poled device is tracked over one week by averaging the relative response variation at all frequencies (e).

## 5. Results and Discussion

Using the methods described in the previous section, we measure the electro-optic coefficients of poled silicon nitride at various modulation frequencies from 50MHz to 15GHz. For the $r_{33}$ component measured with the $TE_0$ mode (Fig.4(a)), the poled device demonstrates an enhancement, up to 30fm/V, across the entire frequency range of testing. The measured $r_{13}$ component (Fig.4(b)) is only enhanced at high modulation speed (>5GHz). The $r_{33}$ value at low frequencies corresponds to a $\chi^{(2)}$ of 0.24pm/V according to [37]

$$r_{ijk} = -2\chi^{(2)}_{ijk} / (\varepsilon_i \varepsilon_j) \qquad (3)$$

where $\varepsilon_i$ is the permittivity of the material. Meanwhile, in both cases, the poled device shows an enhanced high-speed response with a 3dB bandwidth of at least 15GHz, while the non-poled device has a 3dB bandwidth of around 3GHz (we call them the '*fast response*' and the '*slow response*', respectively). As a result, the enhancement of the EO coefficients is most significant in the high-frequency regime (≥5GHz, the blue markers in Fig.4(a) and Fig.4(b)); a >25X enhancement is measured for $r_{33}$ at 7GHz. Note that the EO response of the non-processed device (green markers in Fig.4(a) and Fig.4(b)) is no longer distinguishable from the background noise (typically -130dBm at 1GHz, increasing for higher frequencies) as we increase the modulation frequency to higher than 7GHz and the markers only represent the results when using the noise level as the measurement. Hence, the lower error limits of these points are all set to zero and the upper error limits of the corresponding enhancements are set to infinity. It is reasonable to infer that the actual responses (enhancements) at these frequencies follow the decaying (rising) trend line in the lower frequency regime. We track the induced 15GHz EO response for over one week (red and pink curves in Fig.4(a-b), Fig.4(c)) and observe no significant decay, suggesting that the effect of the poling is long lasting, and very likely permanent.

The improvement of the response speed out of the poling under high temperature confirms that the process induces a second-order nonlinearity in the $Si_3N_4$ core through the alignment of Si-N bonds and breaking of the centrosymmetry, since no other mechanism can provide such a high-speed electro-optic response. We build a physical model to estimate the induced $\chi^{(2)}$ in $Si_3N_4$. The fact that the characterized $r_{13}$ value is smaller than the $r_{33}$ value by a factor of approximately 3 matches well with the analytical result of the model.

We also study the EO coefficients of devices that are processed only with a strong poling electric field or a high temperature (Fig.4(d-e)). No clear change appears after the field is solely applied, yet the heating process reduces the as-deposited EO response. The $r_{33}$ component is suppressed by a factor of 5 while the $r_{13}$ component is completely indistinguishable from the background noise.

The large bandwidth of the measured *fast response* indicates that the RC limit of our device is at least 15GHz, and therefore the *slow response* must originate from a different mechanism of slower speed. Based on the characterization above, we infer that the *slow response* results from carrier effects in the $Si_3N_4$ waveguide due to its positive charging caused by defects [38-39]. The speed limit of the carrier effect electro-optic response is reported to be in the level of 1GHz [40] because of the finite carrier lifetime, which agrees with the measured *slow response* speed. Reported passivation of such defects in $Si_3N_4$ and related effects through annealing at 500°C [41] or higher temperatures [42] also agrees with our observations (Fig.4(d-e), red curves) and supports our conclusions. The deterioration of the EO response after annealing, together with our model, explains the slight decrease of the $r_{13}$ component in the sub-GHz regime after the poling process (Fig.4(b)): the *slow response* is mostly removed in the high temperature while the newly built *fast response* is just not as large in the low-frequency regime. Recent work [29] on the EO response in $Si_3N_4$ devices electrically poled at elevated temperature (260°C, not high enough for defect passivation [42]) likely shares the same carrier-based mechanism. The reported $\chi^{(2)}$ (0.084pm/V) in this work matches the magnitude of the measured *slow response* in our work (10fm/V EO coefficient corresponds to about 0.08pm/V, according to Eq.(3)).

In conclusion, we demonstrate that a long-lasting second-order nonlinearity can be induced in amorphous silicon nitride by poling the material with a strong electric field at high temperature. The induced second-order nonlinearity gives rise to a non-trivial Pockels electro-optic effect in the silicon nitride. This work paves the way to achieve active functions on the $Si_3N_4$ platform. By developing more efficient approaches to pole the silicon nitride film, we estimate it is possible to induce a $\chi^{(2)}$ of 10pm/V or more. With this EO response silicon nitride could become a platform that combines high-speed modulation with low loss, and CMOS compatibility to enable monolithic integration of photonics and CMOS electronics for photonic integrated circuits (PICs), LIDAR, data communications, and quantum optics.

**Funding.** National Science Foundation (ECCS 1941213).

**Acknowledgments.** This work was performed in part at the Cornell NanoScale Facility, a member of the National Nanotechnology Coordinated Infrastructure (NNCI), which is supported by the National Science Foundation (Grant NNCI-2025233). This work was performed in part at the University of Rochester Integrated Nanosystems Center (URnano). This work was performed in part at the Semiconductor and Microsystems Fabrication Laboratory (SMFL) at Rochester Institute of Technology. The authors appreciate the staff at these facilities for their help on equipment during the preparation of the samples.